\documentclass[12pt]{article}
\usepackage{amssymb,amsmath}                                                                                                                                                                                                                                                 \usepackage{openbib}
\usepackage{amsthm}
\usepackage{cite}
\usepackage{url}
\usepackage{color}
\usepackage{graphicx}
\usepackage{bbm}

\oddsidemargin=0cm \theoremstyle{plain}
\newtheorem{lem}{Lemma}
\newtheorem{prop}{Proposition}
\newtheorem{cor}{Corollary}
\newtheorem{theor}{Theorem}

\newtheorem{defn}{Definition}

\def\cf{{\it cf. }}
\def\eg{{e.g.\ }}
\def\ie{i.e.\ }

\def\IX{{\mathbbmss{1}}}
\def\TT{\textsf{T}}

\def\R{{\mathbb{R}}}      

\def\C{{\mathbb{C}}}      
\def\hatw{\widehat}       
\def\H{\mathcal{H}}       
\def\A{\widehat{A}}

\def\Bc{\mathcal{B}}
\def\T{\mathcal{T}}
\def\XX{\mathcal{X}}
\def\B{\widehat B}

\def\F{\mathcal{F}}

\def\S{\widehat{S}}

\def\G{\mathcal{G}}
\def\M{\mathcal{M}}

\def\s{{\it s}}

\def\Re{\mbox {\rm Re} }

\def\tr{\mbox {\rm tr} }

\def\Tr{\mbox {\rm Tr} }

\def\s{\sigma_x}

\def\L2{L^2(\R,\C,dq)}

\def\a{{\bf{a}}}
\def\b{{\bf{b}}}

\def\t{{\bf{t}}}
\def\s{{\bf{s}}}

\def\xx{{\bf{x}}}

\def\yy{{\bf{y}}}

\def\hh{{\it{h}}}

\def\theequation{\arabic{section}.\arabic{equation}}

\begin{document}
\setcounter{equation}{0}\setcounter{lem}{0}

\begin{titlepage}

\font\title=cmbx12 \centerline{{}} \vspace{1cm}
 \centerline{{\title Maximum Bell Violations via Genetic Algorithm Search
}}

\vspace{10mm} \centerline{T. A. Osborn\footnote{E--mail: Thomas.Osborn@umanitoba.ca}, Adam  Rogers\footnote{E--mail: Rogers@physics.umanitoba.ca} } \vskip 8pt \centerline{{\sl
Department of Physics and Astronomy}} \centerline{{\sl University of
Manitoba}} \centerline{{\sl Winnipeg, Manitoba, Canada, R3T 2N2 }}
\vspace{10 mm}

\vspace{10 mm}

  \begin{abstract}\vskip .3 cm

    Bell inequality experiments measure the correlation coefficients of two spatially separated systems.
    In an EPR setup, at one location Alice has  $N_a\geq 2$ observables $A =\{\A_j\}_1^{N_a}$  while at a second remote location Bob has $N_b \geq2 $ observables $B= \{\B_k\}_1^{N_b}$.
    Within this bipartite environment each real $N_a \times N_b$ weight matrix $W$ constructs a Bell operator $\widehat{S}_W$ defined by the $jk$ sum of $W_{jk}\, \A_j \otimes \B_k$.  Operator $\widehat{S}_W$ has the Bell non-locality boundary given by a hidden variable norm of $W$. As the $(A,B)$ composition varies, quantum extremes arise when the $\widehat{S}_W$ operator  norm  has the greatest possible Bell violation.  A genetic algorithm (GA) search over all $(A,B)$ is used to find examples of the Alice and Bob operators that realize quantum extremes. A class $\XX_N$  of weights of special interest is given by the square $N_a=N_b=N$ matrices having two $\pm 1$ entries in each row and column with an odd number of minus signs.
    The class $\XX_N$ is a natural extension of the $2 \times 2 $ CHSH family. For dimensions $N=2\sim10$ the GA search finds that both the EPR correlation matrices and the Bell operator extremes do saturate their respective quantum bounds.
    Maximum Bell operator expectations fall between two benchmarks: the Bell inequality threshold and the quantum bound. The difference between these benchmarks is the quantum gap.
    Weight matrices $W$ that have zero quantum gap are determined by a row, column sum criteria.

   \end{abstract}
    \end{titlepage}

\setcounter{equation}{0}

\section{Introduction}

The boundary between a local classical world and the entangled, non-local quantum realm is defined by Bell inequalities.
Much of the general interest in Bell inequalities \cite{Bell:Physics1964}  is that they encode a form of realism and locality that the Einstein-Podolsky-Rosen (EPR) \cite{PhysRev.47.777} paper asserts is necessary for any consistent, complete theory of physics.
      For many physical observables, Bell inequalities (BI) provide experimentally feasible tests \cite{PhysRevLett.47.460, hensen2015loophole}  for the presence of this locality and its possible quantum violation.\smallskip

 A Bell inequality is not a mathematical inequality but rather a comparison of two conflicting views of nature. Consider a standard two arm EPR setup wherein Alice's wing has $N_a$  observables $\{\A_j\}_1^{N_a}$ and Bob's has $N_b$ observables,  $\{\B_k\}_1^{N_b},\ N_a,N_b \geq 2$. For this system  the general Bell inequality evaluates a weighted combination of pair correlation expectation values $\langle \A_j \otimes \B_k \rangle$ two ways: first via quantum rules and second with theories  consistent with realism and locality.

    In a Bell-EPR experiment, the linear sum of correlations is equivalent to matrix multiplication and so
\begin{equation}\label{EPR1}
  \widehat S_W(A,B) \equiv \sum_{j=1}^{N_a} \sum_{k=1}^{ N_b} W_{jk}\,  (\A_j \otimes \B_k)
\end{equation}
 defines the {\it{quantum Bell operator}} generated by an arbitrary real $N_a\times N_b$ matrix $W$.
  For a system on Hilbert space $\H$ with density matrix $\Omega$,
  the corresponding quantum expectation value is
\begin{equation}\label{EPR1a}
  S_W(A,B|\Omega) = \Tr_\H\, \S_W(A,B)\, \Omega = \sum_{j,k=1}^{N_a,N_b} W_{jk}\, \Tr_\H \big[(\A_j \otimes \B_k)\, \Omega \big] \,.
\end{equation}

   For most choices of the weight matrix
 $W$,  the quantum expectation and the local analog equivalent to   (\ref{EPR1a}) have different numerical ranges. \smallskip

An order $N_a,N_b$  quantum EPR system, with components $(A,B|\Omega)$,  is defined by
\begin{description}
  \item[${\bf{1^0.}}$\ {\it{Bipartite state space }}:]$\H= {{\it{h}}}_a \otimes {{\it{h}}}_b, \quad n_a=\dim {{\it{h}}}_a \geq 2, \quad n_b= \dim {{\it{h}}}_b \geq 2
   $. Alice and Bob Hilbert spaces, ${{\it{h}}}_a$ and ${{\it{h}}}_b$, are generally different, but may be identical.
  \item[${\bf{2^0.}}$\ {\it{EPR observables }}:]  The $\hh_a$ (Alice's) operator set is $A=\{\widehat A_j\}_1^{N_a}$ while the $\hh_b$ (Bob's) set is $B=\{\widehat B_k\}_1^{N_b}$. These observables are self-adjoint.

  \item[${\bf{3^0.}}$\ {\it{Quantum state }}:] The quantum density matrix on $\H$ is a positive trace class operator,
  \ $\Omega >0, \, \Tr_\H\, \Omega =1.$
\end{description}

Many different physical implementations are allowed by these three system requirements. The Hilbert spaces ${{\it{h}}}_a$ and ${{\it{h}}}_b$ are unrestricted. These spaces may be either finite or infinite dimensional. The number of Alice and Bob operators $N_a,N_b \geq2$ can be any finite integer values; the allowed observables in $(A,B)$ are any bounded self-adjoint operators;  the weight matrix $W \in \R^{N_a \times N_b}$ is arbitrary; and, the choice of the quantum state $\Omega$ is open.\smallskip

Bell locality applicable to a $(A,B|\Omega)$ EPR system is realized in terms of a local (or hidden variable) physics picture. This framework, like that of classical statistical mechanics, is based on a classical probability theory whose mean values are computed by integrals over product pairs of variables $a_j,b_k$ (corresponding to the quantum observables $\A_j,\B_k$) with respect to a  probability measure. In this context there is a natural norm that bounds the local Bell expectations.\smallskip

Let $\a$ be an origin rooted, $N_a$ dimensional vector in the hyper-cube  $K_{N_a} = [-1,1]^{N_a}$  and likewise  $\b$ an $N_b$ dimensional vector in $K_{N_b} = [-1,1]^{N_b}$. Denote by $(\,\cdot\, ,\cdot)_{N_a}$ the Euclidean inner product on $\R^{N_a}$.   Introduce a hidden variable based norm for the weight matrix $W$ by
\begin{equation}\label{EPR2}
  ||W||_\star = \max_{\a\, \b} \big|(\a, W \b)_{N_a}\big|
\end{equation}
where the maximum is taken over the $K_{N_a}$ and $K_{N_b}$ hyper-cubes.

     Now suppose Alice and Bob operators are unit operator norm bounded,  $||\A_j||_a\leq 1$, \ $||\B_k||_b \leq 1$. Then all $N_a,N_b \geq 2$ $(A,B|\Omega)$ EPR systems with weight $W \in \R^{N_a \times N_b} $ have a Bell inequality
    \begin{equation}\label{EPR3}
        -||W||_\star \leq S_W(A,B|\Omega) \leq ||W||_\star \,.
    \end{equation}

If $N_a=N_b=2$ and $W_0 = \left(
                                    \begin{array}{cc}
                                      1 & 1 \\
                                      1 & -1 \\
                                    \end{array}
                                  \right)$
then $||W_0||_\star =2$\,  and (\ref{EPR3}) is just the well-known Clauser, Horn, Shimony, Holt (CHSH) inequality \cite{PhysRevLett.23.880}.\smallskip

 As is evident from (\ref{EPR3}), the quantity $||W||_\star$ defines a BI threshold.  Whenever $|S_W(A,B|\Omega)| > ||W||_\star$ the system has a violation of Bell locality.\smallskip

    The  $(A,B|\Omega)$ EPR system has a variety of different, physically significant Bell inequalities. The inequality (\ref{EPR3}) is just one. The essential data acquired in a Bell experiment measures the quantum correlation matrix $C(A,B|\Omega)$ defined by its components
\begin{equation}\label{EPR4}
    C_{jk}(A,B|\Omega) = \Tr_\H \big[(\A_j \otimes \B_k)\, \Omega \big] \qquad j=1, \cdots, N_a\quad  k= 1,\cdots, N_b \,.
\end{equation}
This $N_a \times N_b$ correlation  matrix is real.

The counterpart to the Bell inequalities  are the EPR quantum bounds \cite{Tsirelson1987} which occur when the $(A,B|\Omega)$ configuration allows the greatest possible Bell violation  for a given $W$.
For EPR systems with unit norm bounded $(A,B)$ there are two critical bounds. The  correlation matrix $C(A,B|\Omega)$ has the trace norm bound $\sqrt{N_a\, N_b}$.
The  Bell expectation $|S_W(A,B|\Omega)|$ is bounded by $||W|| \sqrt{N_a \, N_b}$ as one samples all unit norm bounded $(A,B)$ operators and all density matrices $\Omega$.

This paper is organized as follows. Section 2 states and constructs the correlation matrix trace norm bound as well as two Bell operator norm bounds.
  A class of $ \{0,\pm 1\}^{N \times N}$ weight matrices  $W$ with significant overlap with the Bell literature is introduced in the next section. In Section 4,  the genetic algorithm (GA) modeling  of extreme Bell configurations is presented. The quantum gap is the difference between the $W$-dependent quantum extreme  $N||W||$ and the Bell threshold $||W||_\star$. The quantum gap behavior is analyzed  in Section 5.  The class of $N_a \times N_b$ matrices that have vanishing quantum gap is determined.
  Section 6 highlights some of the key numerical EPR features.
      The Bell inequality (\ref{EPR3}) and its generalizations are summarized in Appendix A. \smallskip

Reviews of the Bell inequality literature are found in \cite{clauser1978bell,RevModPhys.65.803,WernerWolf,HGBIX,Banach}.

\section{Quantum Bounds}
\setcounter{equation}{0}

Consider EPR systems having dimensions $N_a,N_b$.
The two basic observables, the correlation matrix $C(A,B|\Omega)$ and the Bell operator $\widehat S_W(A,B)$  have $N_a,N_b$ dependent global bounds as the $(A,B|\Omega)$ composition varies.
Elementary, Banach space arguments establish these bounds and characterize the extreme configurations  $(A^+,B^+|\Omega^+)$ that saturate these bounds. \smallskip

EPR states are constructed from  tensor wave functions $\Psi = \phi_a \otimes \phi_b\,,  \  \phi_a \in \hh_a, \, \phi_b \in \hh_b\, . $
These states are elements of a bipartite Hilbert space $\H = \hh_a \otimes \hh_b$ with an inner product defined by the linear extension of  $(\Psi_1,\Psi_2)_\H = ({\phi_a}_1 , {\phi_a}_2)_a  ({\phi_b}_1 , {\phi_b}_2)_b $ and have norm $||\Psi||_\H = (\Psi,\Psi)^{\frac 12}$.
 The space $\H$ has dimension $n_a n_b \geq 4$.

The analysis of inequalities for  $C(A,B|\Omega)$ and $\widehat S_W(A,B)$ requires a suitable  vector space environment.   This is the following.  Denote the Banach space of bounded operators on $\H$ by $\Bc(\H)$. Let $\Bc_1(\H)\subset \Bc_2(\H)$ \cite{kato1980perturbation} successively be the trace class and the Hilbert--Schmidt class subspaces of $\Bc(\H)$.
Transform $\Bc_2(\H) $  into an EPR compatible Hilbert space, $\H^\sigma$, by adjoining the real inner product $\langle T,S \rangle = \Re \big( \Tr_\H\, T^\dag \, S \big),\ T,S \in \Bc_2(\H)$. The Schmidt norm of $   \Bc_2(\H)$ is ${||T||_\sigma}^2 = \langle T, T \rangle$.  Recall if $\{s_i\}$ are the  singular values of $T$, then $ {||T||_\sigma}^2 = \sum_{s_i > 0} {s_i}^2$. \smallskip

The quantum structure of this system is determined by its density matrix $\Omega$.  This positive operator on $\H$ has a positive square root, $\Omega^{1 \over 2} \in \Bc_2(\H)$. 

 The following lemma shows how to convert the $\A \otimes \B$ quantum expectation into a $\H^\sigma$ inner product.
 The $\hh_a, \hh_b$ operator norms  are $||\cdot||_a$ and $||\cdot||_b$\,; $I_a, I_b$
  are the corresponding identity operators.

\begin{lem} \label{lem1} Suppose the  $\A,\B$ observables are a pair of bounded, self-adjoint operators on $\hh_a, \hh_b$ respectively.  Let $T= (\A\otimes I_b)\, \Omega^{ 1 \over 2}$ and $S = (I_a \otimes \B)\, \Omega^{ 1 \over 2}$. Operators $T,S \in \Bc_2(\H)$ and obey
\begin{equation}\label{L1}
    ||T||_\sigma \leq ||\A||_a \,, \qquad ||S||_\sigma \leq ||\B||_b \,, \qquad
    \langle T, S \rangle = \Tr_\H \big[(\A_j \otimes \B_k)\, \Omega \big] \,.
\end{equation}
\end{lem}
\noindent{{\bf{Proof}.}} The elements $T$ and $S$ are the product of a $\Bc(\H)$ bounded operator and a Schmidt class operator $\Omega^{ 1 \over 2}$, thus $T,S \in \Bc_2(\H)$ and have bounds (\ref{L1}).
The cyclic property of the trace shows
\begin{eqnarray}
\label{L3} \nonumber
  \langle T,S \rangle &=&\Re \big \{ \Tr_\H \big[(\A\otimes I_b)\, \Omega^{ 1 \over 2}\big]^\dag\, (I_a \otimes \B)\, \Omega^{ 1 \over 2} \big \} \vphantom{ \big[ }\\  \label{L4} \nonumber
   &=& \Re \big \{ \Tr_\H\,  \Omega^{ 1 \over 2} (\A\otimes I_b)^\dag\, (I_a \otimes \B) \, \Omega^{ 1 \over 2} \big\} = \Tr_\H \big[(\A_j \otimes \B_k)\, \Omega \big] \,. \qquad \Box
\end{eqnarray}

The relevant collection of Lemma 1 compatible operators for the $N_a,N_b$ $(A,B|\Omega)$ system is  $T_j= (\A_j\otimes I_b) \, \Omega^{1\over 2}, \, S_k = (I_a \otimes \B_k) \, \Omega^{1\over 2}$, $\forall j,k$.
An optimal evaluation of the inner products $\langle T_j, S_k \rangle$ uses a special basis set. Consider the subspace of $\H^\sigma$ defined by closure of the span of the vector family $\{T_j\}_1^{N_a} \cup \{S_k\}_1^{N_b}$. Label this space by $\F(N_a,N_b)$. The number of independent operators in the set $(A,B)$, essentially $\{T_j,S_k| \forall j,k\} $, determines the dimensionality  of this space: $D_{ab} = \dim \big(\F(N_a,N_b)\big) \leq N_a + N_b$. Let $\{e_i\}_1^{D_{ab}}$ be any orthonormal basis of $ \F(N_a,N_b) \subset \H^\sigma$,  $\langle e_i, e_m \rangle = \delta_{im}$. The pair $T,S \in \F(N_a,N_b)$ has the basis expansion

\begin{equation}\label{P4}
    T  = \sum_{i=1}^{D_{ab}} t^i e_i \,, \quad S  = \sum_{i=1}^{D_{ab}} s^i e_i \qquad t^i = \langle e_i, T \rangle\,,
     \quad s^i = \langle  e_i, S \rangle \,.
\end{equation}
In this setting, the $\H^\sigma$ inner product has the Euclidean space equivalent, $\langle T,S\rangle = (\t,\s)$ where $\t = (t^1,t^2,\cdots,t^{D_{ab}}) \in \R^{D_{ab}} $ and $\s = (s^1,s^2,\cdots,s^{D_{ab}}) \in \R^{D_{ab}} $.  Together (\ref{L1})  and (\ref{P4}) imply $||T_j||_\sigma = ||\t_j||$ and $||S_k||_\sigma = ||\s_k||$.

Lemma 1 allows one to write the expectation values entering the quantum correlation matrix via the inner products
\begin{equation}\label{P3}
    C_{jk}(A,B|\Omega) = \Tr_\H \big[(\A_j \otimes \B_k)\, \Omega \big] = \langle T_j,S_k\rangle = (\t_j,\s_k)\,.
\end{equation}

Associate with the $(A,B)$ operator families, the norm means
\begin{equation}\label{P3a}
    \M_a = \bigg( {1 \over {N_a} } \sum_{j=1}^{N_a} {||\A_j||_a}^2 \bigg)^{1/2} \,, \qquad
    \M_b = \bigg( {1 \over {N_b} } \sum_{k=1}^{N_b} {||\B_k||_b}^2 \bigg)^{1/2} \,.
\end{equation} If $||\A_j||_a =1, ||\B_k||_b = 1, \quad \forall j,k$ then $\M_a = \M_b =1$.

\begin{prop} \label{Prop1}  Let $(A,B|\Omega)$ be the configuration of an
   $N_a,N_b \geq 2$ EPR system with bounded operators. The correlation matrix $C(A,B|\Omega)$ has the trace norm bound
\begin{equation}\label{Pr0}
    ||C(A,B|\Omega)||_\tau \leq \sqrt{ N_a N_b}\, \M_a \M_b \,, \qquad \forall \  \Omega\, .
\end{equation}
\end{prop}
\noindent{{\bf{Proof}.}}
With the vectors $\t_j$ and $\s_k$, rendered as columns, define the two matrices
\begin{equation*}\label{G16a}
  {\bf L} = |\t_1|\t_2|\cdots |\t_{N_a}| \in \R^{D_{ab} \times N_a} \,,  \qquad {\bf R} = |\s_1|\s_2| \cdots  |\s_{N_b}| \in \R^{D_{ab} \times N_b}\,.
\end{equation*}
 The $jk$ set of  inner product identities $C_{jk}(A,B|\Omega) = (\t_j, \s_k)$ is equivalent to the matrix factorization
\begin{equation}\label{G16b}
  C(A,B|\Omega)  = {\bf L}(A,B|\Omega)^\TT\, {\bf R}(A,B|\Omega)\,.
\end{equation}
The $\Bc_2(\R^{N_a})$ norm for the matrix ${\bf L}$ is
 \begin{eqnarray}
 \nonumber
    {||{\bf L}||_\sigma}^2  &=& \tr\, {\bf L}^\TT {\bf L} =  ||\t_1||^2 + ||\t_2||^2 + \cdots + ||\t_{N_a}||^2  \\  \label{G16cc}
    &=&  {||\A_1||_a}^2 + {||\A_2||_a}^2 + \cdots +   {||\A_{N_a}||_a}^2 = N_a\, {\M_a}^2\,.
 \end{eqnarray}

 Similarly, the ${\bf R}$ matrix has the Schmidt norm bound ${||{\bf R}||_\sigma}^2 = N_b \, {\M_b}^2$.  These ${\bf L}, {\bf R}$ bounds establish
\begin{equation*} \label{Gxxx}
    ||C(A,B|\Omega) ||_{\tau} = ||{\bf L}^\TT\, {\bf R}||_{\tau} \leq || {\bf L}||_\sigma\, || {\bf R}||_\sigma = \sqrt{N_a\, N_b}\, \M_a\, \M_b \,. \qquad\qquad \Box
\end{equation*}

Based on inequality (\ref{Pr0}) one has a natural definition of a Bell-EPR extreme.

\begin{defn} \label{Df1} An order $N_a,N_b \geq 2$ EPR configuration $(A^+,B^+|\Omega^+)$     is a {{\bf{quantum extreme}}} if
\begin{equation} \label{def1}
   ||C(A^+,B^+|\Omega^+)||_\tau  = \sqrt{N_a N_b}\,  \M_a \M_b\,.
\end{equation}
\end{defn}

 The notation $(A^+,B^+|\Omega^+)$ is reserved for EPR configurations that are quantum extremes.  Definition \ref{Df1} is valid for all dimensions of $\hh_a, \hh_b \geq 2$; including the infinite dimensional cases. Subsequent numerical examples will construct $(A^+,B^+|\Omega^+)$ EPR systems that fulfill the equality (\ref{def1}).

\begin{theor} \label{thm1}  Let $(A,B)$ be an $N_a,N_b \geq 2$ EPR system with bounded operators. The Bell operator ${\widehat S}_W(A,B)$ has the quantum bound
\begin{equation}\label{G19}
||{\widehat S}_W(A,B)||_\H  \leq \sqrt{N_a N_b}\,\M_a \M_b\, ||W||\,, \qquad W \in \R^{N_a \times N_b} \,.
\end{equation}
\end{theor}

\noindent{{\bf{Proof}.}} The correlation matrix $C(A,B|\Omega)$ is trace class for each $\Omega$. Let $\Omega_\Psi = |\Psi\rangle \langle \Psi |$ where $\Psi \in \H$. The $\Psi$-expectation of the hermitian operator ${\widehat S}_W(A,B)$ has the bound
\begin{eqnarray}
\label{G19f}
   |(\Psi, {\widehat S}_W(A,B) \Psi)| &=& \big| \tr\, W^{\TT}\, C(A,B|\Omega_\Psi) \big | \leq  ||W^{\TT}\, C(A,B|\Omega_\Psi)||_\tau \\ \label{G19a}
   & \leq& ||W|| \, ||C(A,B|\Omega_\Psi)||_\tau  \leq \sqrt{N_a\, N_b}\, \M_a \M_b\, ||W||\,. \vphantom {\int}
\end{eqnarray} The first inequality in (\ref{G19a}) follows from  \cite{schatten2013norm} (III. Lemma 8); the final inequality from Proposition \ref{Prop1}. These bounds hold for all unit normed $\Psi$ and establish (\ref{G19}). $\qquad \Box$\smallskip

 Another (and known, \cite{tsirelson1993quantum}) pathway to finding a Bell operator quantum bound is to use Grothendieck's tensor theorem \cite{grothendieck1996resume,pisier2012grothendieck }
which has the following statement:  If $F$ is a real $N \times N$ matrix obeying the restriction
\begin{equation}\label{Gr0}
    |(\xx,F \yy)_N| \leq 1, \ N\geq 2\quad  \text{where}\quad  \xx,\yy \in K_N=[-1,1]^N
\end{equation}
     then there is a constant $K_G(N) > 0$ such that
\begin{equation}\label{Gr1}
    \bigg | \sum_{jk}^N F_{jk}\, \langle X_j, Y_k \rangle_G \bigg | \leq K_G(N) \,.
\end{equation}

Above, $\langle \cdot, \cdot \rangle_G$  is the inner product of  any Hilbert space   $\H_G$ and
$X_j,Y_k$ are any unit bounded elements in this space.
The quantity $K_G(N)$ is Grothendieck's constant of order $N$. Recall, $K_G(N)$ is a bounded increasing function of $N$. Its $N\rightarrow \infty$ asymptotic value satisfies the bound and estimate \cite{krivine1977constante}
 \begin{equation*}\label{G4}
  1.677.. \leq K_G(\infty) \equiv K_G \leq \pi/(2 \ln(1 + \sqrt{2})) = 1.782..
\end{equation*} For small $N$ values,  $K_G(2) = \sqrt{2}\,, \quad K_G(3) < 1.517..\, , \quad K_G(4) \leq \pi/2 \,.$\smallskip

\begin{theor} \label{Tsir} (Grothendieck--Tsirelson). Let $(A,B)$ be an $N_a,N_b \geq 2$ EPR system with unit bounded operators, $||\A_j||_a \leq 1 , ||\B_k||_b \leq 1$.  The Bell operator has the quantum bound
\begin{equation}\label{Gr2}
    || {\widehat S}_W(A,B) ||_\H \leq K_G(N^+)\, ||W||_\star\ , \quad N^+ = \max(N_a,N_b)\,, \quad W \in \R^{N_a \times N_b} \,.
\end{equation}

\end{theor}
 \noindent{{\bf{Proof}.}} First consider the diagonal (symmetric)  problem where $N_a=N_b=N$ and establish that
 \begin{equation}\label{Gr3}
    |S_W(A,B|\Omega)| \leq K_G(N)\,||W||_\star\ ,    \quad \forall \ \Omega \,.
\end{equation}
 The weight matrix $W$ has the HV norm $||W||_\star $. Set\ $ F = W/||W||_\star,\,  ||W||_\star >0$, then requirement (\ref{Gr0}) has the equivalent statement $|(\xx, W \yy)_N| \leq ||W||_\star $\,.  For $\xx, \yy \in K_N$ the definition of $||W||_\star $
ensures that this inequality holds. With this choice of $F$, the second half of the Grothendieck theorem (\ref{Gr1}) is
\begin{equation}\label{Gr4}
    \bigg | \sum_{jk}^N W_{jk} \, \langle X_j, Y_k \rangle_G \bigg | \leq K_G(N)\, ||W||_\star\,.
\end{equation}

Now link $\langle X_j, Y_k \rangle_G$ to  the Bell-EPR framework. Let Hilbert space $\H_G = \H^\sigma$ and set $X_j=T_j, Y_k=S_k$. Lemma 1 shows $\langle T_j, S_k \rangle = \Tr_\H \big[(\A_j \otimes \B_k)\, \Omega \big]$; thereby the left part of (\ref{Gr4}) is $|S_W(A,B|\Omega)|$. This verifies (\ref{Gr3}).
If $\Psi$ is the eigenstate of the Bell operator having eigenvalue $\pm||S_W(A,B)||_\H$ then with $\Omega= |\Psi\rangle \langle \Psi |$ the left side of (\ref{Gr4}) has value $||{\widehat S}_W(A,B)||_\H$.
   So the diagonal version of (\ref{Gr2}) follows.

    The extension to $N_a \neq N_b$ proceeds by enlarging $W$ to a square matrix of dimension $N^+$ and adding suitable $0$ operators to the $(A,B)$ operator set. \quad $\square$ \smallskip

 The Theorem 1 and 2 inequalities  both provide Hilbert space norm bounds for $ {\widehat S}_W(A,B) $. Unlike the correlation matrix bound (\ref{Pr0}), they do not depend on the density matrix $\Omega$. The structure of  Theorem 2 (\cite{Banach}: Theorem 11.12\,) incorporates the Bell threshold $||W||_\star$  in stating the quantum bound.  In contrast, the Theorem 1 quantum bound is independent of $||W||_\star$.

The optimal quantum bound for an $(A,B)$ unit bounded EPR system is the smaller of the Theorem 1\&2 bounds. It is interesting to note \cite{pisier2012grothendieck} that if the Grothendieck constant $K_G$ were $1$ then Theorem 2 predicts that no Bell violations could occur.
  As the weight matrix $W$ varies through the $\R^{N_a \times N_b}$ space,
   the least bound moves between the Theorem 1 and Theorem 2 outcomes. If $W$ is a matrix with a single non-zero  entry $q>0$ then $||W||= ||W||_\star = q$ and the Theorem 2 bound is smaller.  In the $N_a=N_b=N=2$ CHSH case (where $||W_0|| = \sqrt{2},\  ||W_0||_\star = 2 $ ) both Theorem 1 and Theorem 2 have the same upper bound, $2\, \sqrt{2}$.

 The Proposition 1, $C(A,B|\Omega)$ bound does not sense the dimension $D_{ab}$ of  $\F(N_a,N_b)$.  The minimum dimension needed in the inner product representation of the correlation matrix is $N^- = \min(N_a,N_b)$.  This occurs if the vectors $\{\t_j\}_1^{N_a}$ are projected onto the span of $\{\s_k\}_1^{N_b}$ or vice versa.  Using a Clifford algebra representation of operators $\A_j, \B_k$, this type of inner product form of $C(A,B|\Omega)$ was introduced by Tsirelson  \cite{LMP4-93,Tsirelson1987}.

The characterization of all possible EPR correlation matrices is simplified by the Euclidean inner product representation (\ref{P3}).
For $N_a,N_b \geq2$ let $Q(N_a,N_b|\Omega)$ be the set of all correlation matrices having unit norm bounded operators $(A,B)$ and density matrix $\Omega$. Set $Q(N_a,N_b|\Omega)$ is a closed, convex set composed of $\R^{N_a \times N_b}$ matrices.  Within $Q(N_a,N_b|\Omega)$ is a subset of classical correlation matrices $Q^c(N_a,N_b|\Omega)$ selected by the additional condition that they admit a hidden variable representation, \cf (\ref{D1b}).  Although this geometric description of the Bell inequalities has received extensive treatment in the literature \cite{Tsirelson1987,RevModPhys.65.803,landau1987violation,Banach},  it is not essential for the GA investigation of extreme Bell violations.

The quantum bounds in Proposition 1 and Theorems 1\&2 are device independent results that are simple and universal.
They hold for all finite $N_a\geq2, N_b\geq2$. These bounds  do not depend on the dimensions of the Hilbert spaces
$\hh_a,\hh_b$. They are insensitive to the choice of the $\A_j,\B_k$ observables provided that they are norm bounded operators.\smallskip

\section{Extended CHSH Models}\setcounter{equation}{0}
\setcounter{equation}{0}

 For all orders $N_a,N_b$ , the HV inequalities (\ref{EPR3}) says that each weight matrix $W$ generates an associated Bell inequality.   In the symmetrical $(A,B)$ case where $N_a=N_b=N$  we introduce a family of $\{0,\pm1\}^{N \times N}$ weight matrices whose BI's have significant overlap with Bell models found in the literature. These special weight matrices are labeled by $X$ and are referred to as Bell matrices.

\begin{defn} \label{D1} A matrix $X \in \R^{N \times N},\, N\geq2$  is an order $N$  {\bf{Bell matrix}} if
\begin{description}
  \item[$1^0$] Each column and row has 2 non-zero entries with values $\pm 1$.
  \item[$2^0$] $X$ is irreducible.
  \item[$3^0$] $X$ has an odd number of minus signs,  $\nu(X)$.
  \end{description}
\end{defn}

\noindent   The set $\XX_N$ is the collection  of all order $N$  Bell  matrices. \bigskip

\noindent Additional properties of a Bell matrix, $X \in \XX_N$ : \smallskip

 1) The sum of each row (and column) lies in the set $\{-2,0,2\}$;\
 2) All matrices in  $\XX_N$ have the same
 operator norm $||X|| = 2 \cos [ \pi/(2N)]$. \smallskip

\noindent Choosing $W$ to be $X \in \XX_N$ defines the corresponding Bell expectation
\begin{equation}\label{M0}
    S_X(A,B|\Omega) = \sum_{j,k=1}^{N} X_{jk}\, \Tr_\H \big[(\A_j \otimes \B_k)\, \Omega \big] \,.
\end{equation}

The requirement that $X \in  \XX_N $ be irreducible is a natural physical restriction.  If an $X$ were reducible then the resultant BI (\ref{EPR3}) would decompose into two or more disjoint lower order BI's. In this circumstance, the inequality structure would not be a fully coupled order $N$ Bell inequality.

The Bell matrices have nice behavior with respect to row and column manipulations. Denote by $Pr$ any product of signed permutations of the matrix rows; and by $Pc$ any product of signed permutations of the columns.  First, note that the row and column sum restrictions of $X$ to the set $\{2,0,-2\}$ are unchanged by the actions of  $Pr$ and $Pc$. Similarly the number of odd minus signs $\nu(X)$ may change with these transformations, but the number of minus signs remains odd.

For a given $N \geq 2$, the different $X$ matrices are interrelated by unitary transformations. In characterizing these relationships
 a useful special case of $X$  is the following (if $N=4$)
\begin{equation*}\label{C2b}
  Z^0 =  \left(
       \begin{array}{cccc}
         -1 & 1 &0 & 0 \\
         1 & 0 & 1 & 0 \\
         0 & 1 & 0 & 1 \\
         0 & 0 & 1 & 1  \\
       \end{array}
     \right) \ .
\end{equation*}
So $Z^0$ is a tridiagonal Bell matrix with zeros along its interior diagonal and having upper left element $-1$.

The  linkage between different $X's$ of order $N$ takes the following form.
  It is not difficult to show for $N\geq2$ that there exist signed permutation matrices
  $Pr,Pc$ such that
\begin{equation}\label{Ma}
    Z^0 =  Pr X Pc \,, \qquad X \in \XX_N \,.
\end{equation}

    Based on (\ref{Ma}) it follows that $||Z^0|| = ||Pr X Pc|| = ||X||$.  This shows that the
 operator norm of every $X\in \XX_N$ has the same value. The matrix $Z^0$ is hermitian with largest eigenvalue of $2 \cos [ \pi/(2N)]$.  So the common norm value is $||X|| = 2 \cos [ \pi/(2N) ]$.

Obtaining the Bell inequalities for $X \in \XX_N$ requires an estimate of the norm $||X||_\star$\,.

\begin{lem} \label{lem2} Each $X\in \XX_N\,, \ N\geq2$  has the HV norm bound,  $||X||_\star \leq 2(N-1)$.
\end{lem}
\noindent{{\bf{Proof}.}}
Consider the inner product that enters the $||X||_\star$ definition (\ref{EPR2}). Note that $(\a,X\, \b)_N$ is separately linear in $\a \in K_N$ and $\b \in K_N$. As a consequence, the maximum values of the inner product occur on the boundary of the $K_N \times K_N$ support region, where $a_j', b_k' = \pm 1$,
\begin{equation}\label{Mb}
    (\a', X \b')_N = \sum_{j=1}^N  a_j' \left\{  \sum_{k=1}^N  b_k' X_{jk} \right\} \,.
\end{equation}
Applying the Schwartz inequality together with $||X \b'|| \leq ||X||\,  ||\b'||$ gives the bound
\begin{equation*}\label{Mc}
    \big | (\a', X \b')_N \big | \leq || \a'||\, || \b'|| \, ||X|| \leq N \, 2 \cos \left ( \frac \pi{2 N}    \right ) < 2N \,.
\end{equation*}
For each $j$, the inner $k$ sum of (\ref{Mb})  has two non-zero, $\pm 1$ valued terms having summed values $\{0,\pm2\}$. So
 the $K_N$--boundary value of  $\big | (\a', X \b')_N \big |$ must be an even integer less than $2N$, namely $ 2(N-1)$ or smaller.
\qquad \qquad $\Box$ \smallskip

Lemma \ref{lem2} combined with Appendix A: Theorem \ref{HVthm} gives the specialized Bell inequality

\begin{cor}\label{cor2} Let weight $X \in \XX_N$ and  $(A,B|\Omega)$ be an order $N \geq 2$ symmetrical EPR system with unit bounded operators $\A_j,\B_k$. If $HV 1^0$ and Bell locality, $HV 2^0$, hold then
\begin{equation}\label{M1}
    -2(N-1) \leq S_X(A,B|\Omega) \leq 2(N - 1)\,.
\end{equation}
\end{cor}

    The Corollary 1 inequalities readily extend to a larger class of weights than $\XX_N$. The Bell operator is linear with respect to its weight argument,
    \begin{equation*}
        S_{\xi_1 W_1 + \xi_2 W_2}(A,B|\Omega)  = \xi_1 S_{W_1}(A,B|\Omega)  + \xi_2 S_{W_2}(A,B|\Omega)\,, \quad \xi_1,\xi_2 \in \R \,.
    \end{equation*}
Consider the superposition of weights from $\XX_N$: for $\{\xi_i >0\}$ with $\sum_i \xi_i = 1$ and let $W(\xi) = \sum_i \xi_i \, X_i\,, \ X_i \in \chi_N$. Then $S_{W(\xi)}(A,B|\Omega)$ also obeys inequalities (\ref{M1}). \smallskip

For weights $X \in \XX_N$ the Theorem 1 quantum bound reduces to
\begin{equation}\label{D4}
    |S_X(A,B|\Omega)| \leq N ||X|| \leq 2 N \cos [\pi/(2N)] \equiv QB(N) \,.
\end{equation}
For small $N$ values these quantum bounds are $QB(2) = 2 \sqrt{2}\,, \  QB(3) = 3 \sqrt{3}$.
The first of these is Tsirelson's bound for the $N=2$ CHSH system \cite{Tsirelson1987,LMP4-93,landau1987violation,wehner2006tsirelson}. \smallskip

The chained CHSH model of Braunstein and Caves \cite{Braunstein199022} (hereafter BC) defines a particular subset of matrices in $\XX_N$ made from diagonal or anti-diagonal superpositions of the $2 \times 2$ Bell matrices. Within the BC model framework S. Wehner \cite{wehner2006tsirelson}, using semidefinite programming, has established that $QB(N) = 2N \cos [(\pi/(2N)]$ is a sharp quantum bound for the order $N$ Bell expectation ${S}_X(A,B|\Omega)$. In addition, our modeling computations also find that the  Corollary 1 bounds are tight.

 One advantage of the class of weight matrices $\XX_N$ is that the both norms $||X||_\star$ and $||X||$ are simple known functions of $N$.

\section{Maximal Bell Configurations}\setcounter{equation}{0}

 The models in this section explore the EPR behavior of $(A,B)$ bipartite systems that are at or near a  quantum extreme.
Numerical examples test the predictions related to the  ${\widehat S}_W(A,B)$ bounds given in Theorem 1 as well as the Proposition 1 correlation matrix bounds.
Detailed comparisons with the $\XX_N$ Bell inequalities are made.

Using the norm definitions of $||W||$ and $||W||_\star$
it is straightforward to verify the inequality
\begin{equation}\label{lem1a}
    ||W|| \leq ||W||_\star \leq \sqrt{N_a N_b}\,  ||W||\,, \qquad W \in \R^{N_a \times N_b}\,.
\end{equation}

The rightmost inequality in (\ref{lem1a}) ensures that the quantum bound always equals or exceeds the Bell threshold. For order $N$ square matrices $W$ the (\ref{lem1a}) upper bound is sharp and realized by the identity matrix.

\subsection{EPR Quantum extremes}

Profiling quantum extremes has two facets: one focusing on the fundamental Bell operator $\widehat S_W(A,B)$  and a second characterizing the companion correlation matrices $C(A,B|\Omega)$.

 Consider an EPR configuration specified by dimension parameters $(N_a,N_b;n_a,n_b)$ and weight matrix $W \in \R^{N_a \times N_b}$. In a given $\hh_a, \hh_b$ basis,
the $(A,B)$ operator configuration consists of $\A_j
\in \C^{n_a \times n_a}$ and $\B_k \in \C^{n_b \times n_b} $ unit norm bounded hermitian matrices.

A Bell extreme is the supremum of $||\widehat S_W(A,B)||_\H$ as the matrix configuration $(A,B)$ varies over all possibilities consistent with fixed
$(N_a,N_b;n_a,n_b)$.  Denote the relevant set of Alice and Bob operators by
\begin{equation*}\label{Ex1}
  \G(N_a,N_b;\,n_a,n_b) = \G = \big \{ \{\A_j\}_1^{N_a}, \{\B_k\}_1^{N_b} \big | ||\A_j||_a \leq 1, \ ||\B_k||_b \leq 1,\ \forall j,k  \big \}\,.
\end{equation*}
The adjustable variables in $\G$ are the complex entries of $\A_j,\B_k$.
For finite $(N_a,N_b;\, n_a,n_b)$, the set $\G$ is compact and the norm $||\widehat S_W(A,B)||_\H$ is a continuous function on  domain $\G$.  As a consequence there is an $(A^g,B^g)$ element of $\G$ such that
\begin{equation*}\label{Ex2}
  ||\widehat S_W(A^g,B^g)||_\H  = \max_{(A,B) \in \G} ||\widehat S_W(A,B)||_\H. 
\end{equation*}

 An extreme set $(A^g,B^g)  = (\{\A_j^g\}_1^{N_a}, \{\B_k^g\}_1^{N_b})$ is numerically found by using a genetic algorithm \cite{holland1975adaptation,goldberg2002design,rogers2012global} to search the full $\G$ parameter space. This genetic algorithm is the Ferret GA contained in the Qubist Optimization Toolbox for MATLAB \cite{fiege2010qubist}.

   The GA selected extreme ${\G}$-point is not necessarily unique, but the algorithm selects
  a $(A^g,B^g)$  such that
\begin{equation}\label{Ex3}
  ||\widehat S_W(A^g,B^g)||_\H \geq ||\widehat S_W(A,B)||_\H \,, \qquad \forall \ (A,B) \in \G\,.
\end{equation}
So long as the size of the search space $\G$, namely $(N_a\, n_a^2 + N_b\,n_b^2)$, is around 100 or less it is possible to numerically construct matrices $\A_j^g, \B_k^g$ obeying (\ref{Ex3}). In general, the boundary of the space $\G(N_a,N_b;\,n_a,n_b)$ is a multidimensional sector.  As a result a set of GA searches starting with a given $(N_a,N_b,n_a,n_b)$ and $W$ may yield a variety of different  $(A^g,B^g)$ extremes
all having the same Bell operator norm.

One measure of the non-uniqueness of $(A^g,B^g)$ is given by the tensor operator $U_{loc} = U_a \otimes U_b$, with any $\hh_a, \hh_b$ unitary  transformations,  $\A_j''^g = U_a\, \A_j^g U_a^\dag, \ \B_k''^g = U_b\, \B_k^g U_b^\dag $. Operators $\widehat{S}_W(A''^g,B''^g)$ and $\widehat{S}_W(A^g,B^g)$   have the same operator norm.

 Associated with a Bell operator extreme there is a  family of related correlation matrices.
These are obtained from the eigenvalue problem defined by hermitian operator $ \widehat S_W(A^g,B^g)$.   Let $ t \in \T_g = \{1,2,\cdots,\dim \H = n_a\, n_b\}$ index the solutions of the problem
\begin{equation*}\label{Ex4}
    \widehat S_W(A^g,B^g) \, \Psi_t = \lambda_t \, \Psi_t\,, \qquad ||\Psi_t||_\H = 1
\end{equation*}
and order the eigenvalues by $|\lambda_1|\geq |\lambda_2| \geq\cdots \geq |\lambda_{n_a n_b}|$. Define $\T_g^{\max} \subseteq \T_g$ to be the subset of $\T_g$ such that $|\lambda_t| = ||\widehat S_W(A^g,B^g)||_\H$.
The family of projectors $P_t = |\Psi_t \rangle \langle \Psi_t |$ implements the spectral expansion of the Bell operator, $  \widehat S_W(A^g,B^g) = \sum_{t \in \T_g}  \lambda_t\, P_t $.  \smallskip

Each projection $P_t$ defines a correlation matrix,
$ C_{jk}(A^g, B^g| P_t) = ( \Psi_t, \A_j^g\otimes\B_k^g \, \Psi_t)_\H$.
 The eigenvalue $\lambda_t$ records the `size' of the Bell expectation
\begin{equation*}\label{Ex6}
    S_W(A^g,B^g|P_t ) =  (\Psi_t, \widehat S_W(A^g,B^g) \Psi_t )_\H  = \lambda_t \,.
\end{equation*}

The singular values of the  correlation $C(A^g,B^g|P_t)$ determine its trace norm, its Schmidt norm and its operator norm.  Let $S(t)$ be the Schmidt rank of $C(A^g,B^g|P_t)$; and order its singular values: $\mu_1(t)\geq \mu_2(t) \geq \cdots \geq \mu_{S(t)}(t) > 0.$   The correlation matrix norms are
\begin{eqnarray}
\nonumber
  ||C(A^g,B^g|P_t)||_\tau &=& \sum_1^{S(t)} \mu_i(t),  \qquad ||C(A^+,B^+|P_t)||_\sigma^{\, 2}  =  \sum_1^{S(t)} \mu_i(t)^2, \\ \nonumber
  ||C(A^g,B^g|P_t)||_\H  &=&  \mu_1(t)\,, \qquad t \in  \T_g \,.
\end{eqnarray}

If the GA search obtains the Theorem 1 inequality bound, it follows that the companion correlation matrix $C(A^g,B^g|P_t), \ (t\in \T_g^{\max})$ must be a quantum extreme, namely $(A^g,B^g)=(A^+,B^+)$. Specifically,  a saturated outcome for Theorem 1 inequality requires that the supporting correlation matrix be a quantum extreme.

When the GA search finds a $||\widehat S_W(A^g,B^g)||_\H$ norm smaller than $\sqrt{N_a N_b} ||W||$ the numerical model correlations $C(A^g,B^g|P_1)$ obey the Proposition 1 inequality but are not always a Definition 1 quantum extreme.

The various extreme correlation matrices arise from the $\G$-based norm maximum search of $\widehat S_W(A,B)$ for a fixed weight matrix $W$. This process ties together the matrices $W$ and $C(A^g,B^g |P_t)$.

The matrix pair $W$ and $C(A^g,B^g|P_t)$ may be viewed as vectors.   The inner product representation  determines their opening angle to be
\begin{equation*}\label{Ex7}
    S_W(A^g,B^g|P_t) = \langle W, C(A^g,B^g|P_t) \rangle = ||W||_\sigma\, ||C(A^g,B^g|P_t)||_\sigma\, \cos \theta_{Wt}
\end{equation*} specifically
\begin{equation}\label{Ex8}
    |\cos \theta_{Wt} | = \frac { ||\widehat S_W(A^g,B^g) ||_\H} {  ||W||_\sigma\, ||C(A^g,B^g|P_t)||_\sigma } \,, \qquad t \in \T_g \,.
\end{equation}

\subsection{Numerical Bell models}

The numerical models in this work  assume the symmetric EPR case where $N_a=N_b=N$ and in this subsection utilize a Bell matrix weight, $W=X \in \XX_N$.

\begin{figure}
\centerline{\includegraphics[scale=0.9]{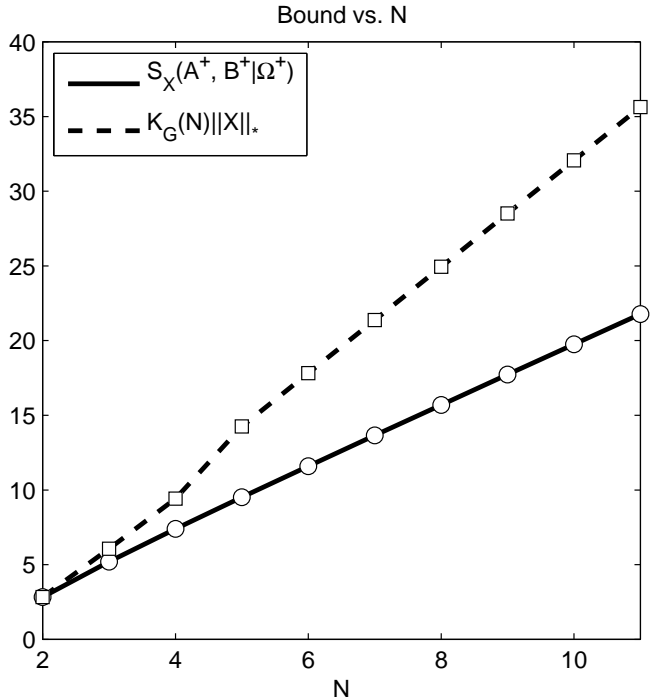}}
\caption{  Theorem $1\&2 $  bounds and the  $|S_X(A^+,B^+|\, \Omega^+ = {P}_{1,2})|$ values. }
\label{fig2}
\end{figure}

First compare  extreme norm values of $S_X(A^g,B^g|P_t),\ (t \in \T_g^{\max}) $ with the $N$ dependent bounds occurring  in Theorems $1\, \& 2$.
Recall that these norm bounds are independent of $n_a,n_b$. For this reason the data points displayed in Fig.~\ref{fig2} have used the lowest dimensional $\hh_a \otimes \hh_b$ realization: $n_a =n_b=2$.
Fig.~\ref{fig2}  plots these extremes as a function of the EPR dimension $N$. The numerical models for $\XX_N$ weight systems show that the Theorem 1 bound, $QB(N)= 2N \cos [\pi/ (2N)] $, is always achieved by an example of $S_X(A^g,B^g|P_t)$.
As Fig.~1 also illustrates, Theorem 1 bound is lower than that of Theorem 2;  $N ||X|| < K_G(N) ||X||_\star $.  For large $N$, $K_G(N) ||X||_\star \approx 1.7 N ||X|| $.

The numerically obtained extreme correlation matrices $C(A^+,B^+|P_t),\  (t \in \T_g^{\max})$ have the following common features. For $X \in \XX_N$ and $N\geq2$ the rank of $C(A^+,B^+|P_t)$ is 2 and the two non-zero singular values are equal, $\mu_1(t) = \mu_2(t) = N/2$. This means that $||C(A^+,B^+| P_t)||_\tau = N, \ (t=1,2)$.  In this way the trace norm bound of Proposition 1 is fulfilled and $C(A^+,B^+| P_{1,2})$ are both quantum extremes as specified by Definition 1. Viewed as vectors in $\R^{N^2}$, the two correlation matrices $t=1,2$ lie on the same line of action with opposite orientation, \ie $C(A^+,B^+| P_2) = -\, C(A^+,B^+| P_1)$. \smallskip

The numerical data associated with the quantum extremes shows a systematic geometry interrelating the 3 matrices, $X \in \XX_N, \ C(A^+,B^+|P_1) , C(A^+,B^+|P_2)$.  Using $||X||_\sigma = \sqrt{2N}, \ ||X|| = 2\, \cos[\pi/(2N))]$ together with  $||C(A^+,B^+|P_{1,2})||_\sigma= N/\sqrt{2}$ and $|S_X(A^+,B^+|P_{1,2})| = N ||X||$ the angle formula (\ref{Ex8}) becomes
\begin{equation}\label{Ex9}
  |\cos \theta_{X\,t} | = \frac {2 \cos  [\pi/(2N)]} {\sqrt{N}}\,,  \quad t = 1,2 \,.
\end{equation} This geometric prediction holds for all finite $n_a \geq 2, n_b \geq 2$.
For increasing values of $N$ the angles are

\begin{equation*}
\begin{array}{ccc}
  N \quad & \theta_{X1} & \theta_{X2} \\
  2 \quad & 0^o & 180^o \\
  3 \quad & 0^o & 180^o \\
  4 \quad & 22.5^o  & 157.5^o \\
  5 \quad & 31.7^o & 148.3^o \\
  \infty  \quad & 90^o & 90^o
\end{array}
\end{equation*}
The zero opening angle examples for $N=2,3$ show that there is a constant $k(N)$ such that
\begin{equation}\label{Ex10}
   C(A^+,B^+|P_{1,2}) = \pm\,  k(N) \, X \,, \quad  N=2,3 \,.
\end{equation} Taking the trace norm of both sides of (\ref{Ex10}) determines the constants to be $k(2) = \sqrt{2}/{2}\,,  \quad k(3) = \sqrt{3}/{2}$.
The numerical modeling data confirms relationships (\ref{Ex10}).  Note also that the angles $\theta_{X\,1,2}$ are independent of $n_a, n_b$.  The numerical data also shows this invariance.  It is to be emphasized that (\ref{Ex10}) holds only for EPR systems with weights in $\XX_2$ or $\XX_3$.

    The outcome in (\ref{Ex10}) means that the $N=2,3$ extreme correlation matrices found from the $\widehat S_X(A^+,B^+)$  supremum norm search can have only one (device independent)  form, namely (with scaling $k(N)$) the Bell matrix, $X$.

\subsection{von Neumann entropy}

The entanglement of a pure state density matrix $\Omega = |\Psi\rangle \langle \Psi|$ , $\Psi \in \hh_a \otimes \hh_b$ is given \cite{barnett2009quantum} by the von-Neumann entropy of the $\hh_a, \hh_b$ reduced density matrices, $\rho_a = \tr _b\, \Omega, \, \rho_b = \tr_a\, \Omega$.  With respect to an orthonormal basis $e_\mu \otimes f_\nu \in \hh_a \otimes \hh_b$ represent this wave function by $\Psi = \sum_{\mu \nu}
J_{\mu\nu}\, e_\mu \otimes f_\nu$. Let $\{s_m\}$ denote the singular values of matrix $J$, then the entanglement of this system is
\begin{equation}\label{Ent1}
    S(\rho_a) = S(\rho_b) = \sum_{s_m > 0} s_m^2 \ln s_m^2 \,.
\end{equation}

\subsection{ Genetic algorithm Bell extremes }

By employing the GA search process, several dozen EPR $(A^g,B^g)$ extremes have been computed for a variety of small values of $(N,n_a,n_b)$. The purpose of this section is to establish that the Theorem \ref{thm1} bounds for $\widehat S_X(A,B)$ are saturated; in the case of the extreme  correlation matrices $C(A^+,B^+|P_{1,2})$ , to reproduce the sum rule bound version of Proposition 1: $\sum \mu_i \leq \sqrt{N_a\, N_b}$. ; and, finally to determine the degree of entanglement of the  $\widehat S_X(A^+,B^+)$ eigenvectors $\Psi_{1}$  and $\Psi_{2}$ which enter the construction of  $C(A^+,B^+|P_{1,2})$.

In all of the GA numerical examples utilizing weights $X \in \XX_N$, the spectrum of the fundamental Bell operator extreme $\widehat S_X(A^+,B^+)$ has a common pattern. The eigenvalues occur in matched $\pm$ pairs. If $\dim \H = n_a\, n_b$ is even the trace is zero; if $\H$ has odd dimension, the trace of equals the one unpaired eigenvalue.

Two representative cases $(N,n_a,n_b) = (3,2,2)$ and $(2,4,3)$ are described in detail.  The predictions of additional  EPR numerical models are summarized in Table 1. \bigskip

\noindent{\bf{Model (3,2,2)}}\quad The weight matrix in $\XX_3$  for this example is $X = \left(
          \begin{array}{ccc}
            -1 & 1 & 0 \\
            1 & 0 & 1 \\
            0 & 1 & 1 \\
          \end{array}
        \right)$.

The norm and quantum gap (\cf Section 5) properties of weight $X$ are: $||X|| = \sqrt{3}$,  $||X||_\star = 4$ and $g(X) = 0.6906.. $

Given $X$, the norm  maximizing GA search of $\widehat{S}_X(A,B)$ finds an extreme Alice and Bob matrix set $(A^+,B^+)$ to be
\begin{eqnarray*}
  A^+_1 &=& \left(
    \begin{array}{cc}
      -0.776327 & -0.052589 -0.628133i \\
      -0.052589 + 0.628133i & 0.776327\\
    \end{array}
  \right) \\
  A^+_2 &=& \left(
              \begin{array}{cc}
                0.0906127 & 0.780399 +0.618682i\\
               0.780399 -0.618682i  &-0.0906127  \\
              \end{array}
            \right)
   \\
   A^+_3 &=&  \left(
              \begin{array}{cc}
               -0.685714 & 0.727810 -0.009451i \\
             0.727810 +0.009451i    &0.685714\\
              \end{array}
            \right)
   \\
  B^+_1 &=& \left(
             \begin{array}{cc}
              -0.984158 & -0.136131 -0.113584i \\
               -0.136131 +0.113584i & 0.984158 \\
             \end{array}
           \right)
   \\
  B^+_2 &=& \left(
             \begin{array}{cc}
              0.497890 &  0.597724 -0.628356i\\
               0.597724 +0.628356i & -0.497890 \\
             \end{array}
           \right)
    \\
   B^+_3 &=& \left(
             \begin{array}{cc}
              -0.486268 &  0.461593 -0.741940i\\
               0.461593 +0.741940i & 0.486268 \\
             \end{array}
           \right)
           \,.
\end{eqnarray*}

In these GA determined EPR matrices, the spectral span for $A^+_j$ and $B^+_k$ is $[-1,1]$; where both endpoints $\pm 1$ are $A^+_j,B^+_k$ eigenvalues.

The quantum bound for $||\widehat{S}_X(A^+,B^+)||$  (Theorem \ref{thm1}, with $\sqrt{N_a \, N_b} =  N=3$) is $6\cos(\pi/6)$. The GA $(A^+,B^+)$ model matrices above closely achieve this bound with a deviation  of $ 3.0 \times 10^{-15}$.

The trace norm  bound (\ref{Pr0}) for the $(3,2,2)$ correlation matrices is $||C(A^+,B^+|P_{1,2})||_\tau =3 $.  This value is reproduced with a deviation of $1.4 \times  10^{-15} $. \smallskip

The rigidity property of the $N=3$ correlation matrix $ C(A^+,B^+|P_1)$ is
\begin{equation*}\label{GA2}
 ||(\sqrt{3}/2)X|| = \frac{3}{2} \,, \qquad  || C(A^+,B^+|P_1) + (\sqrt{3}/2)X|| =  3.5 \times 10^{-8}\ .
\end{equation*}

Finally consider the wave function entanglement associated with this EPR extreme. The tensor expansion of $\Psi_1$ eigenvector of   $\widehat{S}_X(A^+,B^+)$ is realized by a $3 \times 3$ matrix $J$. In the present example this matrix has Schmidt rank $2$.  The probabilities in the open degrees of freedom are $p_1= 1/2$ and $p_2 = 1/2$.  As a result the $\Psi_1$ entanglement is $- \sum p_i\, \ln p_i = 0.693147   $\ .  This is the maximum reduced von-Neumann entropy for a system with two degrees of freedom, namely  $\ln(2)$. The results for the second $\widehat{S}_X(A^+,B^+)$ eigenfunction $\Psi_2$ also have von Neumann entropy $0.693147$.  The 4 correlation matrices constructed from the $4$  eigenvectors of $\widehat S_X(A^+,B^+)$ are all quantum extremes.\smallskip

\noindent{\bf{Model (2,4,3)}} \quad
Here the weight matrix is the CHSH type,  $  X = \left(
        \begin{array}{cc}
          -1 & 1 \\
          1 & 1 \\
        \end{array}
      \right) $.

In this model the norms and quantum gap properties are: $||X||  =\sqrt{2}, \quad ||X||_\star = 2, \quad g(X) = 0.5858.. $

The $(2,4,3)$ quantum bound for $||\widehat{S}_X(A^+,B^+)||$  is $4\cos(\pi/4) = 2 \sqrt{2}$. The GA $(A^+,B^+)$ model matrices achieve this bound with a deviation of $ 2.9\times 10^{-9}$.

The trace norm  bound (\ref{Pr0}) for the two correlation matrices is now $||C(A^+,B^+|{P_{1,2}})||_\tau =2 $.  This singular value sum rule is numerically reproduced with an deviation of $2.0 \times  10^{-9} $.\smallskip

The rigidity property of the $N=2$,  $ C(A^+,B^+|P_1)$ correlation matrices  is also reproduced
\begin{equation*}\label{GA3}
 ||(\sqrt{2}/2)X|| = 1 \,, \qquad  || C(A^+,B^+|P_1) + (\sqrt{2}/2)X|| = 1.5 \times 10^{-5}\,.
\end{equation*}

 An EPR operator configuration $(A,B)$ is called {\it{Clifford}} \cite{Tsirelson1987} if all the $\hh_a$ (and $\hh_b$) anticommutators are proportional to the identity. In Table 1, the extreme models $(N,n_a,n_b)$  are Clifford when their dimensionality $n_a,n_b$ is 2, and not Clifford when $n_a >2 , n_b >2$.  An additional symmetry tied to the $n_a=n_b=2$ case is that the $(A,B)$ operator squares are the identity \ie $(\A_j^+)^2 =I_a, (\B_k^+)^2 =I_b, \ \forall j,k$. This property is not present when $n_a,n_b > 2$.

\begin{table}
\caption{Bell Model Extremes} \vskip .3 cm
\centering
\begin{tabular}{l l l c c c}
\hline\hline
$N, n_a, n_b$ & Thm 1 & Sum Rule & $\Psi_1$ entropy & $\Psi_2$ entropy & quantum extremes \\ [0.5ex]
\hline \\ [-0.5ex]
$2, 3, 2$ & $1.8\  10^{-12}$ & $1.3 \ 10^{-12}   $ & 0.5896$^*$ & 0.5145$^*$ & 2/6 \\
$2, 3, 3$ & $2.4 \ 10^{-9}  $ & $1.7 \ 10^{-9}   $ & 0.6931 & 0.6931 & 2/9 \\
$2, 4, 2$ & $1.6 \ 10^{-7}$ & $1.1 \ 10^{-7}$ & 0.5653$^*$ & 0.5653$^*$  & 4/8\\
$3, 3, 2$ & $2.8 \ 10^{-6} $ & $1.5 \ 10^{-6}$ & 0.5544$^*$ & 0.4583$^*$ & 2/6\\
$3, 3, 3$ & $3.7 \ 10^{-4} $ & $1.4 \ 10^{-4}$ & 0.6932 & 0.6932 & 2/9\\
$3, 4, 3$ & $4.1 \ 10^{-3} $ & $2.1 \ 10^{-3}$ & 0.8765$^*$ & 0.9056$^*$ & 2/12 \\
$4, 2, 2$ & $4.9 \ 10^{-11}$  & $1.4 \ 10^{-11}$ & 0.6931 & 0.6931 & 4/4 \\
$5, 2, 2$ & $5.3 \ 10^{-15}$  & $7.9 \ 10^{-16}$ & 0.6931 & 0.6931 & 4/4
\\ [1ex]
\hline
\end{tabular}
\end{table}

\smallskip

The GA-max searches reported in Table 1 find that the norm $||\widehat S_X(A^+,B^+)||$  saturates the Theorem 1 bound. In these searches the $\A_j,\B_k$ are mutually independent. This common extreme behavior is changed if one or more of the $B$ (or $A$) operator pairs commute.   Consider the case $(N,n_a,n_b) = (3,2,2)$ with $X \in \XX_3$ and let $\B_3$ be a function of $\B_2$.
  With this restricted $(A,B)$ set,  the GA-max Bell expectation $|S_X(A^g,B^g,P_1)| = 4.8284$\, exceeds the Bell threshold 4 but is less than the Theorem 1 bound $N||X||= 5.1963$. This $(3,2,2)$ example has a companion correlation  matrix $C(A^g,B^g|P_1)$ that is a quantum extreme.   If all the $B$ operator pairs commute then the GA-max is at the Bell threshold 4. Lastly,  if all $A$ and all $B$ pairs simultaneously commute then the EPR system has quantum locality (\cf Sect. A.2) with  $||\widehat S_X(A^g,B^g)|| = 4$.
\smallskip

The goal of the GA search in the $(A,B)$ parameter space $\G(N_a,N_b;\,n_a,n_b)$ is to find the global maximum of $||\widehat S_X(A,B)||_\H$ together with the operators $(A^g,B^g)$ that realize this maximum. In practice, the solution space is highly degenerate with many equally valid maxima. Due to the stochastic nature of the GA search, we find that the obtained solutions have a narrow spread
 below and near the global maximum. 
   This is in part due to the finite precision of the elements of the $\A_j, \B_k$ matrix entries, as well as the dimensionality of the search space $\G$. This pattern is evident in the Thm 1 and and Sum Rule  columns of Table 1. The larger deviations from the Sect.~2 bounds are associated with larger search spaces.

\section{Bell Quantum Gap}\label{gap}

\setcounter{equation}{0} 

Consider the $N_a,N_b$ family of Bell operators $\{\widehat S_W(A,B)\}$  (having unit norm bounded Alice and Bob operators) where both the  $(A,B)$ composition and the associated weight $W$ vary. For this system define the {\it{quantum gap}} as the distance between the Theorem \ref{thm1} quantum bound and $W$-dependent Bell threshold,

\begin{equation}\label{g1}
     G(W)  = \sqrt{N_a\, N_b}\,||W|| - ||W||_\star \geq 0 \,.
\end{equation}
This non-negative function is independent of the density matrix $\Omega$ and the $(A,B)$ configuration.
A necessary and sufficient condition for a Bell violation is that $G(W) >0$.\smallskip

This first characterization of the gap (\ref{g1}) has the homogenous property $G(\lambda W) = |\lambda|\,G(W),\ \lambda \in \R$. This is analogous to  the scaling $S_{\lambda W}(A,B|\Omega) = \lambda\, S_W(A,B|\Omega)$.  For this reason $G(W)$ is not useful in comparing gaps for differing $W$.

    A scale invariant quantum gap is
\begin{equation}\label{g2}
    g(W) = \sqrt{N_a N_b} -  \frac { ||W||_\star}  {||W||} \qquad W  \neq 0 \,, \quad N_a,N_b \geq 2 \,.
\end{equation}
From (\ref{lem1a})  it is known that the the smallest value of the ratio $ {||W||_\star} / {||W||} $ is 1. This shows that the biggest possible  scale invariant gap is
\begin{equation*}\label{g3}
    \max_{W \neq \,0}\, g(W) = \sqrt{N_a N_b} - 1\,,  \qquad W \in \R^{N_a \times N_b} \,.
\end{equation*}

The function $\max g(W)$ is  monotonically increasing in the dimensions $N_a,N_b$ with a least EPR value occurring for the CHSH case where $N_a \!=\!N_b\!=\!2$ and $g(W) = 1$. Figure 2 is a plot of the $g(W)$ for various random $3\times 3$  matrices.

\begin{figure} \label{fig1}
\centerline{\includegraphics[bb= 158 248 445 544, clip, scale=1.0]{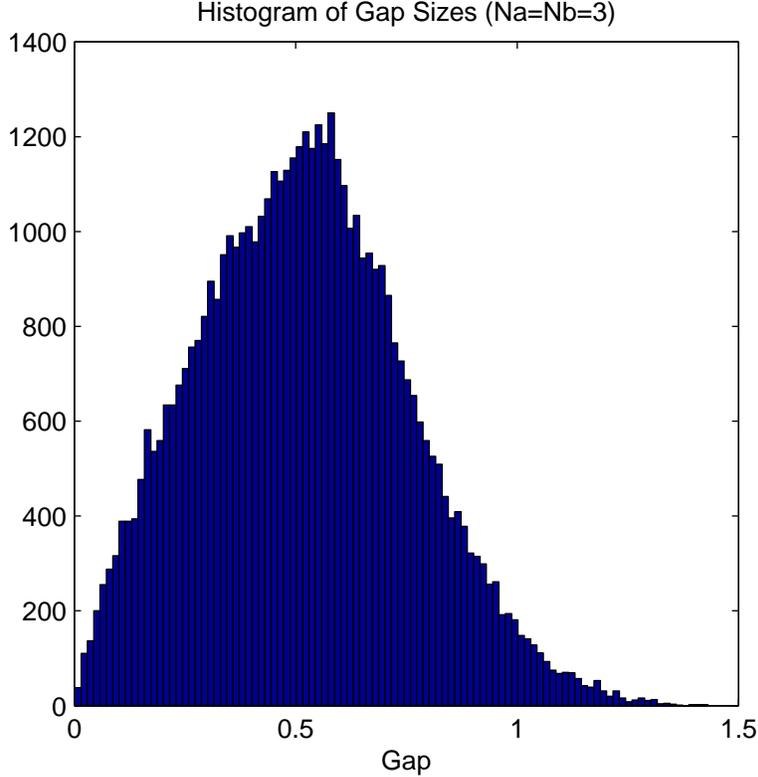}}
\caption{Quantum gap $g(W)$,  $W \in \R^{3 \times 3}$,  sampled over $50,000$ random matrices. }
\label{ls}
\end{figure}

It is of interest to identify the class of weight matrices that have vanishing quantum gap.  The following proposition characterizes this class.  A {\it{signature matrix}}\quad $D$ is a diagonal matrix with $\pm 1$ entries. Note that if $D_1, D_2$ are signature matrices of order $N_a,N_b$ then $||D_1 W D_2||_\star = ||W||_\star$.

\begin{prop} \label{Pr2}  Suppose $W \in \R^{N_a \times N_b}\,, \ N_a,N_b \geq 2$.  The quantum gap $G(W) = 0$ if and only if there are signature matrices $D_1,D_2$ such that $D_1 W D_2$ has all row sums equal to $ \sqrt {N_b/N_a}\, ||W||$ and all column sums equal to $\sqrt{N_a/N_b}\, ||W||$.
\end{prop}
\noindent{{\bf{Proof}.}} Assume that $ G(W)  =0$.
Recall the definition
 of the HV norm is
\begin{equation}\label{T2}
    ||W||_\star = \sup \big \{ |(\a,W\b)_{N_a}|\, : \, \prec \!\a,\b \! \succ\, \in K_{N_a} \times K_{N_b} \big \} \,.
\end{equation}
The map $\prec \!\a,\!\b \succ\, \mapsto |(\a,W\b)_{N_a}|$ is continuous and its domain $K_{N_a} \times K_{N_b}$ is compact so the supremum is realized by a vector pair. Denote one such pair by the tuple $\prec \! \a_w,\b_w \!\succ$,
\begin{equation*}\label{T3}
    ||W||_\star =  \sqrt{N_a\, N_b}\, ||W|| = |(\a_w,W\b_w)_{N_a}| \,.
\end{equation*}

Achieving the supremum in (\ref{T2}) requires simultaneously saturating three inequalities \newline  i)\ $|(\a,W\b)_{N_a}| \leq ||\a|| \, ||W\b||$, ii)\  $||W\b|| \leq ||W||\, ||\b|| $ and iii)\ $||\a|| \leq \sqrt {N_a}, \ ||\b|| \leq \sqrt{N_b}$.
Taken together  the equalities in i) $\sim$ iii) imply the extreme pair $\prec \!\a_w,\b_w\! \succ$ obeys
\begin{equation}\label{T5}
    W \b_w = c\, \sqrt \frac {N_b}{N_a}\, ||W||\, \a_w\,,  \qquad  W^{\TT} \a_w  = c\, \sqrt \frac {N_a}{N_b}\, ||W||\, \b_w\, \,,
\end{equation} where $c=\pm 1$.  The fact that $W^{\TT} W > 0$ means that the $\pm$  sign choice of $c$ in these two identities is the same.  The conditions $||\a_w||=\sqrt{N_a},\, ||\b_w|| = \sqrt{N_b}$ indicate that the $\a_w,\b_w$ vectors have end points located at the corners of $K_{N_a}$ and $K_{N_b}$, respectively.

    Next, reconfigure the information in (\ref{T5}) to make the row and column content explicit. The effect of a signature matrix is to transform one $K_{N_a}$ corner vector into another. Let $D_1(\a)$ be the signature matrix with diagonal elements $a_j= \pm 1, j=1,..,N_a $. Denote by $\IX_a$ the all $+1's$ corner vector in $K_{N_a}$. Then $\a =D_1(\a) \IX_a$. Likewise let $D_2(\b)$ be the signature matrix defined by the corner vector $\b \in K_{N_b}$, then $\b = D_2(\b) \IX_b$.

    Define $M(\a_w,\b_w) = D_1(\a_w) W D_2(\b_w)$.  In terms of these signature matrices, (\ref{T5}) becomes
    \begin{equation}\label{T6}
        M(\a_w,\b_w) \IX_b = c\, \sqrt \frac {N_b}{N_a}\, ||W||\, \IX_a\,, \qquad M(\a_w,\b_w) ^\TT \IX_a =c\,  \sqrt \frac {N_a}{N_b}\, ||W||\, \IX_b \,.
    \end{equation}
The left sides of these equations are the row and column sums of $M(\a_w,\b_w)$.  If the sign of $c$ is $-1$, changing the overall sign of one of the signature matrices ensures positive row column sums. The construction above shows that if $W$ has 0 quantum gap then there exists signature matrices $D_1(\a_w), D_2(\b_w)$ that such $M(\a_w,\b_w)$ has row and column sums with the constant values given in (\ref{T6}).

The only if statement in Proposition \ref{Pr2} results as follows.  Here it is assumed that the identities in (\ref{T6}) hold for a $K_{N_a} \times K_{N_b}$ corner vector pair say, $ \prec \! \a' , \b' \! \succ$, where these vectors replace $ \prec \! \a_w, \! \b_w \succ$ in (\ref{T5}).
  Norm $||W||_\star$ has the lower bound
\begin{eqnarray*}
\nonumber
  ||W||_\star &\geq&   |(\a', W \b')_{N_a}| = |(D_1(\a') \IX_a, W D_2(\b') \IX_b)_{N_a}| \vphantom{ \sum }\\
   &=& |(\IX_a, D_1(\a') W D_2(\b') \IX_b)_{N_a}| = \sqrt{N_a\, N_b} \, ||W||\,.
\end{eqnarray*} However the HV norm also has the upper bound is $||W||_\star \leq \sqrt{N_a N_b}\, ||W||$. Thus $ ||W||_\star = \sqrt{N_a\, N_b} \, ||W||$ whenever (\ref{T6}) holds for any corner pair.  \qquad  \qquad $\Box$ \smallskip

    There are many zero gap matrices in $\R^{N_a \times N_b}$. If $\prec \!\a, \b \!\succ \, \in K_{N_a} \times K_{N_b}$ is any pair of corner vectors, then the outer product $W= |\a \rangle \langle \b|$  has $G(W)=0$.  These matrices have rank 1.  But higher rank vanishing gap matrices also exist \cf (\ref{magic}).

The Theorem 2 inequality also defines a possible Bell violation window based on the Grothendieck constant.  In this latter case the allowed gap is $||W||_\star(K_G(N^+) -1)$. The Theorem 2 characterization of the quantum bound environment is deficient in that it does not predict a zero gap weight matrix.

\subsection{Zero gap examples}

In the case of $0$-gap weight matrices $W$ (of dimension $N$) one continues to have quantum extremes but in this context the norm extremes for $\widehat{S}_W(A^+,B^+)$, namely $N||W||$ (Theorem 1) and $||W||_\star$ (Bell threshold) are the same.

An EPR model that illustrates this special  behavior is the following.  Consider a system of dimension $(N,n_a,n_b) = (3,3,3)$ with a weight matrix

\begin{equation} \label{magic}
 {W_m} =
\left(
  \begin{array}{ccc}
    8 & 3 & 4 \\
    1 & 5 & 9 \\
    6 & 7 & 2 \\
  \end{array}
\right)\ .
\end{equation}
This dimension 3 (magic square) matrix has 0-gap with norms $||{W_m}||_\star = 45$, $||{W_m}|| = 15$.\smallskip

The GA norm search constructs $\widehat{S}_{{W_m}}(A^+,B^+)$.  Among the $9$ spectral values of this operator only one state $\Psi_1$ has eigenvalue equal to $||\widehat{S}_{{W_m}}(A^+,B^+)|| = ||{W_m}||_\star = 45$.
The related $\Psi_1$ correlation matrix is extreme with all elements $C_{jk}(A^+,B^+|P_1) = 1, \ \forall \, jk$.

The trace norm for $C(A^+,B^+|P_1)$ is 3. The numerical model gives this value to within machine precision.

The striking feature of this $(A^+,B^+)$ configuration  is that the $\widehat{S}_{{W_m}}(A^+,B^+)\,\Psi_1 = 45\, \Psi_1$  wave function is not entangled; its entropy is zero with no significant error.  Finding extreme configurations that have an $\widehat{S}_{W}(A^+,B^+)$  eigenvector with no entanglement is a rare occurrence.

A closely related model continues to use the magic weight ${W_m}$ but has different $(A,B)$ dimensions: (3,3,2).
The GA determined norm extreme again obeys  $||\widehat{S}_{{W_m}}(A^+,B^+)||= 3||{W_m}||=45$ with a saturated sum rule $||\widehat{S}_{{W_m}}(A^+,B^+)||_\tau =3$. These results hold to machine accuracy.
The spectrum of $\widehat{S}_{{W_m}}(A^+,B^+)$ has a sign-flipped pair of large eigenvalues $\lambda_1= -\lambda_2 = ||\widehat{S}_{{W_m}}(A^+,B^+)||$.  In contrast to the (3,3,3) model, the associated eigenstates $\Psi_1$ and $\Psi_2$ both have finite entropy, $0.486\ $.

The GA-found extremes in both the models (3,3,3) and (3,3,2) have the feature that $\A_j = \A_r, j\neq r$ and $\B_k = \B_s, k \neq s$. So this pair of configurations are quantum local (\cf Sect. A.2) in that all pairwise $A$ and $B$ commutators vanish.

 A relevant instance of a Bell operator consistent with Proposition 2 occurs in  Heydari  \cite{Heydari}. For an order $N = 4 $ EPR system let $\{\A_j\}_1^4 , \{\B_k\}_1^4$ be dichotomous operators, \ie ${\A_j}^2 = I_a, \ {\B_k}^2 = I_b$ and choose the weight matrix to be
\begin{equation*}
    W_{00} = W_0 \otimes W_0 = \left(
                                 \begin{array}{cccc}
                                   1& 1 & 1 & 1 \\
                                   1 & -1& 1 & -1\\
                                   1& 1 & -1& -1\\
                                   1 & -1 & -1 & 1 \\
                                 \end{array}
                               \right)\, .
\end{equation*}

 Investigation \cite{Heydari} aims to estimate the Grothendieck constant $K_G(4)$ (\,and subsequently $K_G(N)$ for larger $N$)  by  calculating  the maximum Bell violation for this EPR configuration  under the assumption that the best quantum bound is the Theorem \ref{Tsir} inequality. However $W_{00}$ is a 0-gap matrix and so no Bell violation is possible.  In detail, the Bell threshold is $||W_{00}||_\star =8$; Theorem 1 bound is $4 \times ||W_{00}|| = 8$ while the  Theorem 2 bound is $8\, K_G(4) \approx 8\, \pi/{2} = 12.56..$\,.
    The signature matrices required by Proposition 2 are $D_1 = D_2 = diag({\bf{d}}), \ {\bf{d}} = (1,1,1,-1)$.

\section{Model Conclusions}

The Bell-EPR systems with unit bounded operators have numerical quantum extremes with the following key properties:

\begin{description}
  \item[i)] For EPR systems with Bell matrix weights $X \in \XX_N \ (N=2\sim 10)$ the extremes for $||\widehat{S}_X(A,B)||$ saturate the quantum bound, $N ||X||$. When $N=2$, the Theorem 1 bound equals the Theorem 2 bound. If $N > 2$ the Theorem 1 bound is smaller than the Grothendieck bound, \cf Fig. 1.
  \item[ii)] For $X \in \XX_N$ with $N=2,3$ the extreme correlation matrices have a rigid structure:  $C(A^+,B^+|P_{1,2})= \pm k(N) X, \ \, k(2) = \sqrt{2}/{2}\,,  \ k(3) = \sqrt{3}/{2}\,$.
  \item[iii)] If the Bob (or Alice) operators are restrained, \eg $\widehat{B}_3 = f(\widehat{B}_2)$ for some function $f$, then the $(N,n_a,n_b)=(3,2,2)$  GA max $||\widehat{S}_X(A^g,B^g)||$ still violates the BI but does not reach the $N||X||$ bound. Nevertheless, the associated correlation matrix $C(A^g,B^g|P_1)$  is a quantum extreme.
  \item[iv)] For arbitrary weight matrices $W$, the quantum gap vanishes when the quantum bound is equal to the Bell threshold, $||W||_\star$\,. A row, column sum criterion (Proposition 2) defines the zero gap weight matrices. The $3\times3$ magic matrix is a zero gap matrix.
\end{description}
\begin{appendix}
\section{Bell Locality}
\setcounter{equation}{0}
\renewcommand{\theequation}{\Alph{section}.\arabic{equation}}

This Appendix recounts  a traditional version of the Bell inequalities that matches the style and structure of the Section 2 quantum bounds and which makes explicit the locality foundations of the BI.

A local theory or hidden variable picture defines a generic classical framework wherein the physical observables are represented by functions (random variables) on a classical probability space. In detail, there is a triple $(\Lambda, {\cal F},\mu )$ composed of a sample space, $\Lambda \subseteq \R^n, n\geq 1$, together with an integration theory suitable for HV state averaging.  Let ${\cal F}$ be a Borel $\sigma$-algebra of subsets of $\Lambda$, and $\mu: {\cal F} \rightarrow [0,1]$ a positive, unit normalized probability measure.

The quantum triplet analogous to
$(\Lambda, {\cal F},\mu )$ is $(\H,\Xi,\Omega)$ where $\H$ is the system Hilbert space, $\Xi$ denotes a collection of observables appearing in the Bell inequality of interest and $\Omega$ is the system density matrix. The operator family $\Xi$\, includes all of the Alice and Bob observables,  $\widehat A_j, \widehat B_k$\,.

Each observation in the HV picture has two elements. First, the system state is characterized  by a hidden variable
$\lambda \in \Lambda$. For the observable $R$ (corresponding to the quantum $\widehat R$\,)  the system status in state $\lambda$ is given by a random variable $r\!:\!\Lambda \rightarrow \R$ with value $r(\lambda)$. Second, the HV mean is the integral of $r$ with respect to the probability measure $\mu$. In the usual HV context, the individual values $r(\lambda)$ are not available; only the mean values are known.  In the case of several random variables $r,s,t, \cdots$ all the HV means are determined by this one common (averaging) measure $\mu$ which expresses the statistical distribution of HV states incident on Alice and Bob's measuring instruments.

    The characterization of the system state by $\lambda$ together with observable properties $r(\lambda), s(\lambda), t(\lambda), \cdots$ implements the realism aspect of the hidden variable theory. These properties are understood to be always present and independent of any observation.

This HV summary retains the locality and realism interpretations of the hidden variable picture and emphasizes how the norm $||\cdot||_\star$ naturally arises from this environment.

\vskip -.4 cm

\subsection{HV axioms}

\noindent HV\,$1^0$ (Spectrum rule)\  The random variable $r:\Lambda \rightarrow \R$ corresponding to the quantum observable $\widehat R \in \Xi$ has  values $r(\lambda)$ restricted to the spectral span of $\widehat R$.
\\[1mm]

\noindent HV\,$2^0$ \,(Bell locality) For each product pair $\widehat R\, \widehat S$ of commuting, self-adjoint operators in $\Xi$,  the quantum expectation and the HV mean  agree
\begin{equation}\label{HVPR}
    \Tr_\H \widehat R\, \widehat S\, \Omega  =  \int_\Lambda  r(\lambda) s(\lambda) \, d \mu(\lambda)  \equiv \langle R\,S \rangle_{HV}
\end{equation}
or in abbreviated form, $\langle \widehat R\widehat S \rangle_{QM} = \langle RS \rangle_{HV}$.\smallskip

  The commutative, numerical multiplication $r(\lambda) s(\lambda) = s(\lambda) r(\lambda)$ algebraically implements the Bell locality process.
The rules above make explicit the  assumptions built into  Bell's original work \cite{Bell:Physics1964}.
  The notation and framework in HV\,$ 1^0, 2^0$  is adapted from the work of Fine \cite{fine1982hidden,fine1982joint}
   and Malley \cite{PhysRevA.58.812}. \smallskip

In the EPR setup, the $\widehat R, \widehat S$ commutativity arises because $\A_j, \B_k$ are operators acting in different Hilbert spaces. In addition, the locations of Alice and Bob usually are placed sufficiently far apart so that their respective measurements are space-like separated.  This supports the view (incorporated into HV $1^0, 2^0$) that the random variables $a_j(\lambda)$ and $b_k(\lambda)$ are mutually independent and insensitive to the experimental measurement of each other.   The locality statement HV $2^0$, equivalently (\ref{HVPR}), is the core foundation of the Bell inequality.\smallskip

Non-commutativity in the EPR framework resides between operators in set $A$ on Hilbert space $\hh_a$ \eg $[\A_j,\A_m] \neq 0\,, \ j\neq m$;
 likewise for the operator set $B$ on space $\hh_b$.

Within the HV theory framework, an order $N_a,N_b$  EPR system corresponding to the quantum configuration $(A,B|\Omega)$ is labeled by $(a,b|\mu)$.  It is defined by the properties
\begin{description}
  \item[H1.\ {\it{A sample (state) space}}:] $\Lambda\subseteq \R^n, n\geq 1$ whose elements $\lambda$ are HV states.

\item[H2.\ {\it{HV observables}}: ] The $\{a_j\}_1^{N_a}, \{b_k\}_1^{N_b}$ (associated with $\{\A_j\}_1^{N_a} \,
    $and$ \, \{\B_k\}_1^{N_b}$)  are random variables with respect to measure $\mu$. Let $a_j^-$, $a_j^+$ ($b_k^-$, $b_k^+$) denote the least and greatest eigenvalues of $\A_j$ (and $\B_k$).   As a consequence of HV $1^0$, $a_j(\lambda) \in [a_j^-,a_j^+], \ b_k(\lambda) \in [b_k^-,b_k^+],\, \lambda \in \Lambda$.
 \item[H3.\ {\it{A system state}}:] This is a unit normalized probability measure $\mu$ on $\Lambda$, $\int_\Lambda d\mu=1$.
\end{description}

 The expectation for the configuration $(a,b|\mu)$ having weight matrix $W$ is the $\lambda$ integral
\begin{eqnarray} \label{D1a}
  S_W(a,b|\mu)  &=& \int_\Lambda \big(\a(\lambda), W \b(\lambda)\big)_{N_a} \, d\mu(\lambda) \\ \label{D2}
  \big(\a(\lambda), W \b(\lambda) \big)_{N_a} &=& \sum_{j=1}^{N_a} \sum_{k=1}^{N_b} a_j(\lambda) W_{jk}\, b_k(\lambda) \,.
\end{eqnarray}
The quantity $S_W(a,b|\mu)$ is the HV analog of the quantum expectation $S_W(A,B|\Omega)$.\smallskip

Applying Bell locality, HV $2^0$, to the quantum correlation components gives
\begin{equation}\label{D1b}
    C_{jk}(A,B|\Omega) = \int_\Lambda a_j(\lambda)b_k(\lambda) \, d\mu(\lambda) \equiv c_{jk}(a,b|\mu)\,, \qquad \forall j,k \,.
\end{equation} Here $c_{jk}(a,b|\mu)$ is the HV mean.  Combined (\ref{HVPR}) and (\ref{D1b}) imply
\begin{equation}\label{D1c}
    S_W(A,B|\Omega) = S_W(a,b|\mu)\,.
\end{equation}

The widest version of the BI accepts arbitrary bounded $\A_j,\B_k$ without requiring that these operators be unit bounded. In this case, the norm construct $||\cdot||_\star$ needs to be adjusted to fully sample the spectrums of Alice and Bob's observables as given by axiom HV $1^0$.   This enlarged vector norm \cite{horn2012matrix} on matrices  is
\begin{equation*}\label{G1}
    ||W||_{A,B} = \max \big \{ |\big(\a, W \b \big)_{{N_a}}|:   a_j \in [a_j^-, a_j^+],\,  b_k \in [b_k^-,b_k^+], \forall j,k \big \} \,.
\end{equation*}

An operator valued, density matrix independent Bell inequality is given by\vskip .2 cm

 \begin{theor}\label{HVthm} (Bell).  {\it {Suppose $\widehat{S}_W(A,B)$ is a Bell operator of order $N_a,N_b \geq 2$ having bounded $\A_j,\B_k$ operators and weight $W \in \R^{N_a \times N_b}$. If $HV 1^0$ and Bell locality, HV $2^0$,  hold then} }
 \begin{equation}\label{H1}
     || \widehat{S}_W(A,B)||_\H \leq ||W||_{A,B}\,.
\end{equation}
\end{theor}
\noindent{\it{\bf{Proof}}}. Let the EPR system have density matrix $\Omega$.  The $||\cdot||_{A,B}$ norm provides $\lambda$-independent bounds of the inner product integrand in (\ref{D1a}), thus
\begin{equation}\label{F1}
 \big |S_W(a,b|\mu) \big| \leq \int_\Lambda \big |\big(\a(\lambda), W \b(\lambda)\big )_{{N_a}} \big|\, d\mu(\lambda) \leq \int_\Lambda ||W||_{A,B}\, d\mu(\lambda) = ||W||_{A,B} \,.
\end{equation} Bound (\ref{F1}) together with the Bell locality property (\ref{D1c})  establishes
\begin{equation}\label{F1a}
  |S_W(A,B|\Omega)| \leq ||W||_{A,B} \,.
\end{equation}
Revise (\ref{F1a}) by setting $\Omega = |\Psi\rangle \langle \Psi| $ where $\Psi$ is any unit normed $\H$ state. This implies (\ref{H1}).
 $\ \Box$ \smallskip

  Whenever $(A,B)$ are unit norm bounded, then $||W||_{A,B} = ||W||_\star$ and (\ref{H1}) is consistent with (\ref{EPR3}). The Bell inequality (\ref{H1}) is universal  in the sense that it applies to all $N_a,N_b$ EPR systems with $\A_j,\B_k$ bounded operators,  all Hilbert space dimensions $n_a,n_b \geq 2$, all weight matrices $W \in \R^{N_a \times N_b}$, and all density matrices $\Omega$.

The simple HV averaging analysis establishing the inequality (\ref{F1}) for $S_W(a,b|\mu)$  is Bell's original argument \cite{Bell:Physics1964} realized in terms of the norms $||\cdot||_{A,B}$ (\,or $||\cdot||_\star$).
   To obtain the final BI (\ref{H1}) one needs only HV axiom $2^0$. In this regard Theorem \ref{HVthm}   has used minimal assumptions about the structure of the local HV world and how it couples to quantum mechanics.

\subsection{Quantum Locality}

Whenever the $(A,B)$ operator sets are fully commutative then the Born quantum expectation (\ref{EPR1a}) admits a local probability representation similar to the HV form (\ref{D1a}).

Let the set of $N_a$ Alice's operators mutually commute. Then there is \cite{plesner1969spectral} an $\hh_a$-operator $\hatw E = \hatw E^\dag$ commuting with all of Alice's operators and a set of real measurable functions $\{\widetilde{a}_j\}_1^{N_a}$ such that
$\A_j = \int_{\R_a} \widetilde{a}_j(\lambda_a) \, d\epsilon(\lambda_a)$
where $\epsilon$ is the spectral measure of $\hatw E$ and $-||A_j||\leq \widetilde{a}_j(\lambda_a) \leq ||A_j||$.  Similarly for the pairwise commuting bounded operator set $\{\B_k\}_1^{N_b}$ there is a family of functions $ \{ \widetilde{b}_k \}_1^{N_b}$ such that $\B_k = \int_{\R_b} \widetilde{b}_k(\lambda_b) \, d\tau(\lambda_b)  $
 where $\tau$ is the spectral measure of an $\hh_b$-operator $\hatw D = \hatw D ^\dag$ that commutes with all of Bob's observables and $-||B_k|| \leq \widetilde{b}_k(\lambda_b) \leq ||B_k||$.  The spectral expansion of the Bell operator $\hatw S_W(A,B)$ EPR has the tensor form
\begin{eqnarray}
 \nonumber  {\hatw S}_W(A,B) &=& \sum_{jk}\, W_{jk} \int_{\R_a} \widetilde{a}_j(\lambda_a)\, d\epsilon(\lambda_a) \otimes \int_{\R_b} \widetilde{b}_k(\lambda_b) \, d\tau(\lambda_b)  \\ \nonumber
   &=& \int_{\R_a \times \R_b} \big( \widetilde{a}(\lambda_a)\,, W \widetilde{b}(\lambda_b)\big)_{N_a} \,
    d\epsilon(\lambda_a) \otimes d\tau(\lambda_b) \leq \ ||W||_{A,B}\,  I_\H\,.
\end{eqnarray}
This operator bound implies that $| S_W(A,B|\Omega)| \leq  ||W||_{A,B}$ for all density matrices on $\H$.
So Bell violations are not possible in the quantum local EPR configuration.\smallskip

\end{appendix}\vskip .3 cm

\noindent{\bf{Acknowledgments}}\vskip .2cm

The authors thank our colleagues  K-P. Marzlin, M. Kondratieva,  F. Molzahn, J. Fiege, and I. Cameron for critical and helpful  discussions as this work developed. Additionally, we thank S. Kirkland for the insights that led to the proof of Proposition 2.

\end{document}